\documentclass[aps,twocolumn,prb,superscriptaddress,amsmath,tightenlines]{revtex4}

\usepackage{graphicx}
\usepackage{epsfig}

\newcommand{\ra}{\rangle}
\newcommand{\la}{\langle}

\begin{document}

\title{Quantum Annealing Machine based on Floating Gate Array}
\author{T. Tanamoto}

\author{Y. Higashi}

\author{T. Marukame}
\affiliation{Corporate R \& D center, Toshiba Corporation,
Saiwai-ku, Kawasaki 212-8582, Japan}

\author{J. Deguchi}
\affiliation{Center for Semiconductor R \& D, Toshiba Corporation,
Saiwai-ku, Kawasaki  212-8520, Japan} 

\begin{abstract}
Quantum annealing machines based on superconducting qubits, which have 
the potential to solve optimization problems faster than digital computers,
are of great interest not only to researchers but also to the general public. 
Here, we propose a quantum annealing machine based on a semiconductor floating gate (FG) array. 
We use the same device structure as that of the commercial FG NAND flash memory 
except for small differences such as thinner tunneling barrier. 
We theoretically derive an Ising Hamiltonian from the FG system in its single-electron region. 
Recent high-density NAND flash memories are subject to intrinsic obstacles that originate from their small FG cells. 
In order to store information reliably, the number of electrons in each FG cell should be sufficiently large. 
However, the number of electrons stored in each FG cell becomes smaller and can be countable.
So we utilize the countable electron region to operate single-electron effects of FG cells.
Second, in the conventional NAND flash memory, the high density of FG cells induces the problem 
of cell-to-cell interference through their mutual capacitive couplings.
This interference problem is usually solved by various methods using software of error-correcting codes. 
We derive the Ising interaction from this natural capacitive coupling.  
Considering the size of the cell, 10 nm, the operation temperature of quantum annealing processes is expected 
to be approximately that of a liquid nitrogen. 
If a commercial 64 Gbit NAND flash memory is used, ideally we expect it to be possible to construct 
2 megabytes (MB) entangled qubits by using the conventional fabrication processes in the same factory 
as is used for manufacture of NAND flash memory.
A qubit system of highest density will be obtained as a natural extension 
of the miniaturization of commonly used memories in society.
\end{abstract}

\maketitle
\section{Introduction}
Artificial intelligence (AI), whose progress is a momentous trend in 
science and technology, is expected to lead to drastic changes in society.
Faster solving of combinatorial optimization problems is a prerequisite for 
efficient development of AI algorithms such as machine learning algorithms.  
A quantum annealing machine (QAM) is expected to solve the combinatorial
optimization algorithms of NP-hard problems in a shorter time than 
is possible with classical annealing methods. 
Nishimori {\it et al.} developed the theoretical foundation of the QAM
\cite{Nishimori,Finnila}, and QAMs based on superconducting circuits are 
widely used~\cite{Dwave,Dwave2}.

NP-hard problems such as the traveling salesman problem
can be mapped to the problems to find ground states of the Ising Hamiltonian,
 expressed by~\cite{Lucas}
\begin{equation}
H=\sum_{i<j} J_{ij} s_i^z s_j^z 
+ \sum_i h_i s_i^z, 
\end{equation}
where the variable $s_i$ is a classical bit of two values($s_i=\pm 1$).
The first term is the interaction term with a coupling constant $J_{ij}$, 
and the second term is the Zeeman energy with an applied magnetic field $h_i$.
This classical Ising machine has already been realized by a conventional semiconductor memory\cite{Yamaoka}.
In the case of a quantum annealing machine, a tunneling term is added and expressed by
\begin{equation}
H=\sum_{i<j} J_{ij} \sigma_i^z \sigma_j^z 
+ \sum_i [h_i \sigma_i^z + \Delta_i (t) \sigma_i^x]
\label{QAM}
\end{equation}
where the variables are expressed by Pauli matrices $\sigma_i^\alpha$ ($\alpha=x,z$) instead of digital bits. 
The tunneling term is controlled such that it disappears at the end of the calculation, given by
\begin{equation}
\Delta(t\rightarrow \infty) \rightarrow 0.
\end{equation}
The Hamiltonian (\ref{QAM}) can be found in many physical systems in nature. 
To use physical systems as QAMs,  the tunneling term and the Ising term should be 
controlled separately and locally by electric gates. 
The development of QAMs based on superconducting qubits have advanced the furthest~\cite{Dwave}. 
The advantage of superconducting qubits lies in the long coherence time of superconducting states.

Let us consider the possibility of constructing QAMs based on semiconductor devices.
The greatest advantage of using semiconductor devices is that 
the smallest artificial structures at the highest density can be manufactured in factories.
Obviously, the competitor of the QAM is the conventional digital computer. 
For the QAM to prevail, 
integration of a huge amount of cell is unavoidable. 
During  the past 20 years, many qubits based on semiconductor devices have been investigated~\cite{Ladd}.
In quantum computation, accurate control of wave functions is required 
from initial states to final states for measurements.
Because it is not easy to precisely control the wave functions of electrons and spins in semiconductor devices, 
even two-qubit logic operation has been proven only recently~\cite{Siqubit}.
Moreover, it is known that the coherence times of charge qubits 
are shorter than those of spin qubits. 
On the other hand, in the quantum annealing machine, 
the  condition of the strict control of wave functions can be loosened 
provided that the final state is an eigenfunction of the target Hamiltonian, 
and the intermediate processes can include disturbance with several kinds of noise.
Thus, we can exploit the advantages of semiconductor devices 
such as high integration and productivity, for a QAM.

Here, we propose a QAM using the conventional NAND flash memory
consisting of  floating gate (FG) cells~\cite{Masuoka,Samsung,Toshiba}.
NAND flash memory has a dominant share of the growing market for storage applications
extending from mobile phones to data storage devices in data centers~\cite{Takeuchi}. 
The NAND flash memory has the advantages of high-density memory capacity and low production cost per bit with low power consumption and high-speed programming and erasing mechanisms.
Data storage of personal computers is also transitioning from hard disk to flash memory.
An FG cell corresponds to 1 bit for a single level cell type and $m$ bit for a multi-level type. 
Each FG is typically made of highly doped polysilicon and placed in the middle of a gate insulator of a transistor~\cite{Brown,Aritome}. 
If there is no extra charge in an FG, the cell behaves like a normal transistor. 
In the programming or writing step, electrons are injected into the FG by 
applying voltage to the control gate. In the erasing step, electrons are 
ejected from the FG to the substrate by applying voltage to the back gate.
The amount of the charge of FG determines the threshold voltage above which the 
current between the source and drain changes. 
In NAND flash memories, the FG cells are connected like a NAND gate.
In general, the distance between the FGs is of the same order as 
the size of FG, resulting in high-density memory.
For example, Sako {\it et al}. developed 64 Gbit NAND flash memory
in 15 nm CMOS technology~\cite{Toshiba}, which 
is organized by a unit of 16KB bit-lines $\times$ 128 word-lines.
This means that the number of closely arrayed FG cells in a single unit is 16KB$\times$128$\approx$ 2MB.
The integration and miniaturization of flash memory cells have progressed continuously and 
the current flash memories have stacked 3D structures using trapping layers~\cite{Samsung2,Toshiba2}.

We theoretically prove that a two-dimensional (2D)  FG array can be used as a QAM.
The QAM proposed here has the structure shown in Fig.1.
The FG cells are capacitively connected to each other, which is the same arrangement as that in a commercial flash memory.
The fundamental idea is that we will be able to regard a small FG cell in the single-electron region as a charge qubit. 
Now the size of the FG NAND flash memory is 15 nm~\cite{iedm2012,iedm2013}, 
and the size can shrink to beyond 7 nm\cite{TSMC7nm,Samsung7nm,IBM7nm}. 
In the flash memory with 15 nm cell,  single-electron effects 
can be observed at room temperature~\cite{Nicosia}.
When the doping concentration of electrons is $10^{20}$ cm${}^{-3}$, 
the number of electrons in a volume of 15$\times$15$\times$50 nm${}^{3}$
is about 1125 and countable.
Once we can control the single-electron effects, we can realize a two-level 
system by using a crossover region between two different quantum states 
with different numbers of electrons, following such as Ref.~\cite{Makhlin}.

\begin{figure}[h]
\centering
\includegraphics[width=8.0cm]{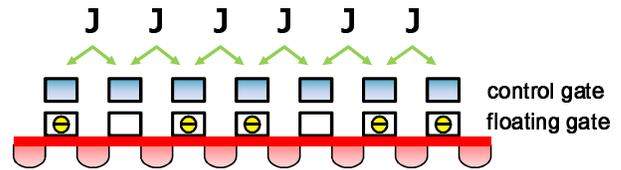}
\caption{
A block of a proposed quantum annealing machine based on a floating gate (FG) array 
has the same structure as commercial NAND flash memory other than 
the thickness of tunneling barrier.
A 2D FG memory array is constructed by arranging many blocks on the substrate.   
The interference effect between the FG cells is the origin of the Ising interaction.
In order to use bidirectional tunneling between FG and substrate, 
we simultaneously apply voltages to both the gate and substrate.
Because we can use the same process as for conventional memory, the production cost is low 
and the fabrication process is well established.
}
\label{fig1}
\end{figure}
The distance between FG cells is of the same order as the size of FG cells in conventional 
NAND flash memories.
Thus, the capacitive coupling cannot be neglected~\cite{Lee}.
The changes of electrostatic potentials of neighboring cells affect the 
electronic static potentials of target FG cells. 
In regard to the conventional use of the memory cell, 
these capacitance couplings between FG cells are undesirable. 
In the conventional NAND flash memories, this effect is alleviated by computer software 
of various error-correcting codes. 
But, here we use this coupling as the interaction between quantized cells.
We derive the Ising interaction term and the Zeeman term in Eq.~(\ref{QAM}) from a simple capacitance network model. 

In the conventional usage of FG memory, 
the gate voltage is applied either positively or negatively corresponding to 
WRITE (programming data) and ERASE (erasing data) processes, respectively.
Thus, in order to realize the tunneling between the FG cells
and the substrate in both directions, 
both gate electrode and substrate have the same voltage 
whereas the voltage of FG should be low. 
By controlling the voltages, we realize the third term of Eq.~(\ref{QAM}). 
 
This QAM has two major structural differences compared with conventional FG memories. 
The first major difference is that the
tunnel barrier between FGs and substrate in the present proposal is thinner than 
that of the commercial FG array, in order to realize the bidirectional tunneling.
The second major difference is that we do not use the air-gap between FG cells~\cite{iedm2013}
in order to increase the electrostatic coupling between FG cells.

Semiconductor devices consist of many different regions on the substrate.
There are several doping concentrations with n-type and p-type semiconductors.
The conventional FG devices are based on the silicon transistor with additional 
floating gates. Thus, there are several regions with different electrostatic potentials.
It is necessary to investigate how the above-mentioned idea can be implemented 
in a realistic potential profile.
Here, we use a technology computer aided design (TCAD) tool to
check whether we can generate the desired potential profile of tunneling within an appropriate voltage region.  
Although this simulator does not include the single-electron effect and we provide results of coupling three cells, we can 
sketch a rough engineering drawing in a realistic situation.

The great advantage of using semiconductor devices instead of superconducting devices 
is that they are the smallest structures that humans can reliably integrate in Gbit order. 
Our proposed QAM will be able to be fabricated using a mask set 
and a product line that are almost same as those used for the conventional NAND flash memory.
We believe that integration capabilities backed by mass-production technologies 
will eventually lead to quantum computers that are far more controllable than those available today.
This QAM will be of far more benefit than QAMs without commercial antecedents 
in mass-productions,
because the progress from fundamental science to a commercial product 
involves a huge cost and time.


This paper is organized as follows:
In Sec.~\ref{sec_optimization}, we briefly review an implementation of a QAM 
for a 2D qubit array.
In Sec.~\ref{sec_idea}, we explain the basic idea of how to implement QAM into an FG array.
In Sec.~\ref{sec:form}, we derive an Ising Hamiltonian 
from an FG array.
In Sec.~\ref{sec_tcad}, we present the results of TCAD simulation based on realistic FG sizes.
In Sec.~\ref{sec_decoherence}, we discuss the noise and decoherence problem of our QAM.
In Sec.~\ref{sec:discuss}, we discuss the possibility of 
high-frequency operation. 
We close with a summary and conclusions in
Sec.~\ref{sec:conclusion}. 
Additional information is presented in the Appendix.
\section{How to construct QAM in 2D NAND flash memory structure}\label{sec_optimization}
The simpler the structure and mechanism, the greater is the reduction in variations among cells, and the more stable the operations become. 
Thus, in order to realize a reliable large-scale QAM,  the structure and mechanism should 
be as simple as possible.  
For the simplest QAM, the coupling $J_{ij}$ has a fixed value determined by its 
cell structure.
The typical optimization problem with a fixed $J_{ij}$ is the MAX-CUT problem~\cite{MAXCUT,Kahruman}.
The MAX-CUT problem is given as follows: 
Given a graph $G(V,E)$, $|V | = n$ with weights $J_{ij}$ for
all $(i,j) \in E$, 
the max-cut is a problem of 
finding $S \subseteq V$ such that the cut
$S, V \backslash S$ is the maximum.
Introducing the variables:
$s_i =1$ if $i \in S$, $s_i =-1$ if $i \in V \backslash S$.
The max cut problem can be expressed as
finding the minimum of the Ising model of 
$\sum_{i<J} J_{ij} (s_is_j-1)$ under a constraint 
of $s_i^2=1$ for  $i=1,...,n$.
The MAX-CUT problem~\cite{MAXCUT,Kahruman} is applied to the noise reduction 
problem~\cite{Yamaoka}.
On the other hand, in general, the values of the coupling $J_{ij}$ are chosen 
depending on each optimization problem~\cite{Lucas}. 
For example, in the traveling salesman problem, 
$J_{ij}$s are taken as the distances between two cities~\cite{Lucas}.
Thus, there is a trade-off between the range of application and the simplicity of device structure. 
Here, we consider a QAM whose coupling $J_{ij}$ is determined by 
the capacitive coupling between two FG cells.
Because the distances between FG cells are all the same, 
a fixed value of $J_{ij}$ is provided in our system in its original condition.
Note that this limitation can be resolved and 
arbitrary variable of $J_{ij}$ can be realized by 
applying high-frequency electric fields as shown in Ref.~\cite{Niskanen}.
Because we would like to present the fundamental proposal of QAM
based on FG arrays,
we will concentrate on our QAM for a conventional operation region of the fixed $J_{ij}$.

In addition, we consider a 2D planer FG array. 
In order to solve general problems in the QAM,  all connections between two cells are required~\cite{IonTrap}.
This means that the coupling $J_{ij}$ should be formed in any two cells.
A coherent Ising machine~\cite{CIM,Yamamoto} can realize this situation by using optical systems.
Because the connections between solid-state cells are fixed, 
it is not possible to directly connect two arbitrary cells in 
solid-state circuits with minimum distances.  
If we apply the methods proposed in Ref.~\cite{Valiant,Choi,Lidar},
we can implement connections between distant qubits.
We can also use the advantage of conventional circuits, {\it i.e.}, 
we can implement digital circuits for connecting distant FG cells 
similar to Ref.~\cite{Yamaoka}.  In this case, 
there is no quantum coherence between digitally connected FG cells.
Although the speed up is less than that of a fully quantum-connected QAM,
it is still expected to operate in the partial quantum regions.

Let us think about using
the lithography mask set of the conventional NAND flash memory of Ref.~\cite{Toshiba}.
The `page' in NAND flash memory corresponds to the number of the digital bits in the bit-line. 
Because Ref.~\cite{Toshiba} has 2 bits per cell, the number of connected FG cells 
in the direction is 128 and one block consists of 16KB $\times$ 128=16384 $\approx$ 2MB cells.
These blocks are digitally connected and form the single plane of the memory.
This means that the coupling $J_{ij}$s exists between nearest neighboring 2MB cells.
Thus,  ideally, 2MB FG charge qubits form an entangled state.
By connecting the blocks digitally, we would be able to 
realize a quantum annealing machine like that of Ref.~\cite{Yamaoka}.
Hereafter, we would like to prove that the 2D FG cell array 
realizes the Ising Hamiltonian with the fixed $J_{ij}$. 
The digital connection and concrete application to annealing problems are beyond the scope of this paper.

\section{Basic idea of QAM based on FG array}\label{sec_idea}
We would like to conceptually explain how to realize a QAM in the conventional FG structure.
\subsection{Floating gate cell}
\begin{figure}
\centering
\includegraphics[width=3.0cm]{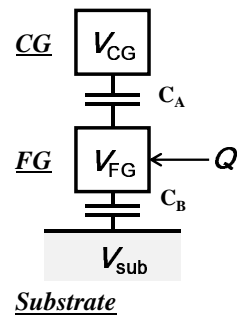}
\caption{
A single cell of the floating gate (FG) array. Charges are stored in the FG.
The charges $Q$ and the potential of the FG, $V_{\rm FG}$, are
controlled by the control gate (CG) voltage, $V_{\rm CG}$ 
and the substrate voltage $V_{\rm sub}$.
}
\label{fig2}
\end{figure}

First, let us consider the single-electron effect in FG cells. 
The basic physics of the FG cell can be explained by the simple model 
of series of capacitance depicted in Fig.~\ref{fig2} given by~\cite{Pavan}
\begin{equation}
Q=C_B(V_{\rm FG}-V_{\rm sub})-C_A (V_{\rm CG}-V_{\rm FG}),
\label{eq1}
\end{equation}
where $Q$ is a charge stored in the FG. $V_{\rm CG}$, $V_{\rm FG}$ and $V_{\rm sub}$ 
are potential energies of the control gate(CG), FG, and substrate, respectively.  
From Eq.~(\ref{eq1}), we have
\begin{equation}
V_{\rm FG}=\frac{Q}{C_A+C_B} + \frac{C_A V_{\rm CG}+C_B V_{\rm sub}}{C_A+C_B}.
\label{simple_V}
\end{equation}
This equation shows that the FG potential energy $V_{\rm FG}$ is determined by 
the charging energy of FG (1st term of the right side of the equation) and 
the controllability of CG and substrate.
When the capacitance $C_{\rm eff}\equiv C_A+C_B$ is sufficiently small, 
the change of the number of single electrons $Q\pm e$ can be 
detected if $e/C_{\rm eff}$ is larger than the operation temperature $T$.
If we simply express $C_{\rm eff}=\epsilon_{\rm ox}S/d_{\rm ox}$, 
where $\epsilon_{\rm ox}$, $S$, and $d_{\rm ox}$ 
are the dielectric constant of a tunneling oxide SiO${}_2$, 
the area, and the thickness of the capacitor, 
and use the value of $\epsilon_{\rm ox}=3.9 \times 8.854 \times 10^{-21}$ F/nm, 
$S=15 \times 15$ nm${}^2$, and $d_{\rm ox}$=3.5 nm, 
we have the voltage shift, $e/C_{\rm eff}$=0.00721 eV =83.8 K, 
by the change of a single electron~\cite{Nicosia}. 
(The Boltzmann constant $k_B =8.617 \times 10^{-5}$ K/(eV) is used.)
Thus, the single-electron effect will be detectable.
Because the single-electron region can 
be used to construct a two-level system as Makhlin {\it et al.}~\cite{Makhlin} showed, 
the small size of the FG gate array allows construction of  
a two-level system, which is quantitatively discussed in the next section.

As in the field of FG memory, we define a coupling ratio $CR$ defined by 
\begin{equation}
CR=\frac{C_A}{C_A+C_B},
\label{CR}
\end{equation}
as an indication of the degree of  controllability of the gate electrode.
We also assume that the capacitance is represented by 
a simple parallel plate capacitor such as
\begin{equation}
C_{\rm A}=\epsilon_{\rm A} LW /d_A, \ \ C_{\rm B}=\epsilon_{\rm ox} LW /d_{\rm ox}, 
\end{equation}
where $L$ and $W$ are the length and  width of the FG cell, 
$d_A$ and $d_{\rm ox}$ are the thickness of tunneling oxides, 
and $\epsilon_{\rm A}$ and $\epsilon_{\rm ox}$ are dielectric constants.
Then, when $d_{\rm ox}$ and $CR$ are given, $d_A$ is given by 
$CR$ and $d_A$ is given by
\begin{equation}
d_A=\frac{1.0-CR}{CR}\frac{\epsilon_A}{\epsilon_{\rm ox}}d_{\rm ox}.
\end{equation}

\subsection{Interaction between FG cells}
In commercial 2D NAND flash memories~\cite{Toshiba,Samsung,iedm2013,iedm2012} 
the distance between FG cells is of the same order as the size of the FG cells.
Thus, the interference effects between FG cells are one of the major issues 
in NAND flash memories~\cite{Lee}.  In order to reduce the interference 
between FG cells, the air-gap technologies are used~\cite{iedm2013}, 
because the dielectric constant of air is smaller than that of the tunneling oxide 
such as SiO${}_2$ (whose dielectric constant is 3.8). However, here we 
use SiO${}_2$ between FG cells to increase the electrostatic coupling between FG cells.

\subsection{Tunneling}
\begin{figure}[h]
\centering
\includegraphics[width=7.0cm]{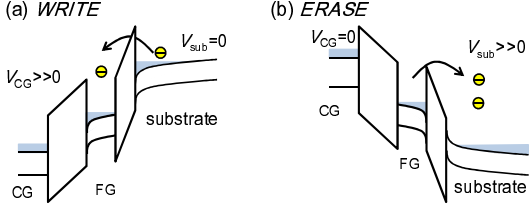}
\caption{
Conventional process of writing data ((a)) and erasing data ((b)).
In the conventional FG array,  the tunneling oxide 
is thick ($\sim$ 7 nm), and a large voltage ($\sim 20$ V) is 
applied to carry out both WRITE and ERASE processes.
}
\label{fig3}
\end{figure}

\begin{figure}
\centering
\includegraphics[width=6.0cm]{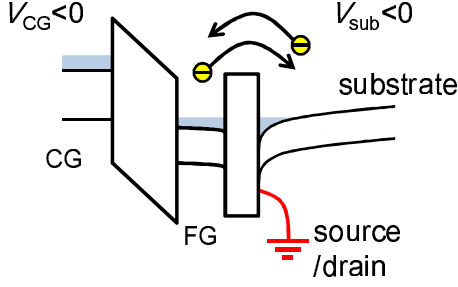}
\caption{
Proposed idea for both bidirectional electron tunneling.
The voltages of the gate and substrate are raised 
compared with the voltages of the source and drain.
The tunneling rate is controlled by the relative potential 
between the gate/substrate and the source/drain. 
}
\label{figConcept}
\end{figure}

In the conventional FG memory, tunneling phenomenon of electrons is used for 
WRITE and ERASE processes of data by biasing the voltage 
between the gate and substrate electrodes. 
However, in the conventional WRITE and ERASE processes, 
electron flow is unidirectional as shown in Figs.~\ref{fig3}(a) and (b).
In the case of a QAM, electron tunneling should be
bidirectional between the FG and substrate.
Bidirectional tunneling occurs when we lower the potential of the 
interface between the FG and substrate.
This situation will be realized by biasing the gate and substrate electrodes 
compared with the source and drain as shown in Fig.\ref{figConcept}.

In the conventional FG memory, the thickness of the tunneling oxide is about 7 nm to store 
charges for more than a couple of years when there is no bias on the gate and substrate.
For the WRITE or ERASE process, a large bias of more than 10 V is applied 
to add or extract electrons from FGs. In this biasing process, because of the 
large applied bias, the tunneling potential changes and Fowler-Nordheim 
tunneling is realized~\cite{Pavan}. 
In order to realize bidirectional tunneling, we consider a case of thin tunneling oxide.

\subsection{FG array}
\begin{figure}
\centering
\includegraphics[width=8cm]{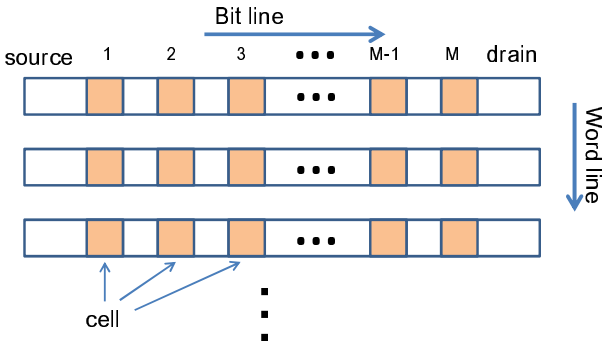}
\caption{
The conventional FG array can be used for the QAM.
The source and drain are attached to both sides of the FG array in the direction of bit-lines.
}
\label{figarray}
\end{figure}

Figure \ref{figarray} shows a schematic of a 2D FG cell array.
The FG cells are coupled capacitively with each other in both word and bit directions.
The source and drain are attached to the both ends of bit-lines. 
When we look at each bit-line, 
biases are applied to the source and drain as well as to the gate. 
Thus, the regions between FG cells in the bit-line 
are in electronically floating states. 
In order to switch on the tunneling of all FG cells at once, 
we will have to apply higher gate voltage to the cells far from the source and drain.

\section{Derivation of QAM Hamiltonian from FG array}\label{sec:form}
Here, we derive the Ising Hamiltonian from the FG array described in Fig.~1
by using a capacitance network model (Fig.~\ref{fig4}).
The charging energy is reduced to the 1st and 2nd term of Eq.~(\ref{QAM}). 
The last term of Eq.~(\ref{QAM}) is derived from tunneling 
of carriers between FGs and substrate.
We assume that the FG cells are sufficiently small that 
the single-electron effects is observable.
The Ising interaction is derived from the interference effect 
between cells~\cite{Lee}. This interference effect is 
the result of Coulomb interaction between the charges of neighboring FG cells.
Usually this interference effect is an obstacle to the conventional flash memories.
But here we can use it effectively for the cell-to-cell interaction.

\subsection{Ising interaction parts and Zeeman energy}
\begin{figure}[h]
\centering
\includegraphics[width=8.0cm]{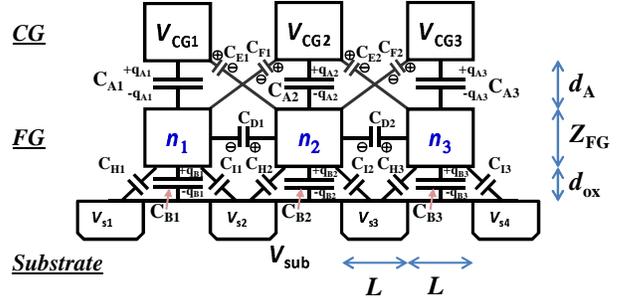}
\caption{
The capacitance network model of three floating gate (FG) cells.
The electronic states are controlled by 
the control gate (CG), substrate, source and drain voltages. 
The width (or depth) of each FG cell is assumed to equal its length, $L$, and 
the distance between two FG cells also equals $L$.
 }
\label{fig4}
\end{figure}

Here, we derive the 1st and 2nd term of Eq.(\ref{QAM}) 
by describing the FG cells using the capacitance network model described in Fig.~\ref{fig4}.
\subsubsection{Analytical formation of Ising interaction parts and Zeeman energy}
The charging energy of $M$ FG cells is expressed by
\begin{eqnarray}
\lefteqn{ U=\frac{1}{2} 
\sum_{i=1}^M \left[
 \frac{q_{A_i}^2}{C_{A_i}}  +\frac{q_{B_i}^2}{C_{B_i}}+\frac{q_{D_i}^2}{C_{D_i}}
+\frac{q_{E_i}^2}{C_{E_i}} +\frac{q_{F_i}^2}{C_{F_i}}
+\frac{q_{H_i}^2}{C_{H_i}}
+\frac{q_{I_i}^2}{C_{I_i}} 
\right]   }
\nonumber \\ 
&-&
\sum_{i=1}^M [
(q_{A_i}+q_{E_i}+q_{F_{i-1}})V_{{\rm CG}i} +q_{B_i}V_{\rm sub}+q_{H_i}V_{s_i}+q_{I_i}V_{d_i} ],
\nonumber \\
\label{UU0}
\end{eqnarray}
where $q_{E_{M+1}}=q_{F_{M+1}}=0$, and $V_{d_i}=V_{s_{i+1}}$.
The number of charges $n_i$ in the $i$-th FG is given by
\begin{equation}
n_i=-q_{A_i} \!+q_{B_i}\!-q_{D_{i}}\!+q_{D_{i-1}}\!-q_{E_{i-1}}
\!-q_{F_i}\!+q_{H_i}\!+q_{I_i},\!
\end{equation}
where $q_{D_0}=q_{E_0}=0$.
The capacitances are defined by 
\begin{eqnarray}
C_{\rm D}&=&\epsilon_{\rm ox} Z_{\rm FG}W /X_{\rm D}, \ 
C_{\rm E}=\epsilon_{\rm ox} (LW/2) /X_{\rm E},  \\ 
C_{\rm H}&=&\epsilon_{\rm ox} (LW/2)/X_{\rm H},
\end{eqnarray} 
with 
$X_{\rm E}=\sqrt{(L/2)^2+d_A^2}$ and
$X_{\rm H}=\sqrt{(L/2)^2+d_{\rm ox}^2}$. 
For simplicity, from now on we formulate a case of three cells.
\begin{figure}[h]
\centering
\includegraphics[width=5.5cm]{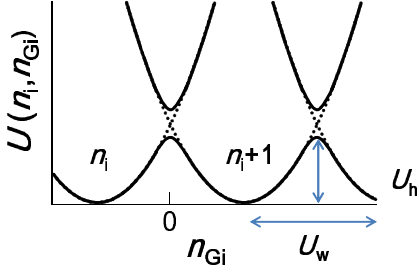}
\caption{
A schematic of the charging energy of a single FG as a function of $n_{G_i}$, which 
represents a gate voltage of an $i$-th cell
as shown in Ref.~\cite{Makhlin}.
The crossover region $n_{G_i}\approx 0$ of two parabolas
realizes a two-state quantum system.
$U_h$ indicates a typical scale by which the single-electron effect can be observed or not.
$U_h$ indicates a typical scale by which the single-electron effect can be controlled or not.
See Fig.~\ref{figUhw}.
 }
\label{fig5}
\end{figure}
The charge distribution is obtained after minimizing the capacitance energy 
by using the method of Lagrange multipliers~\cite{tana0,tana1} and given by
\begin{eqnarray}
\lefteqn{ U = \sum_{i=1}^2 \left[ 
 \frac{1}{2C_{a_i}} 
\left(1+\frac{C_{D_i}^2}{C_{a_{i}}C_{a_{i+1}}} \right) (n_i+Q_{v_i}^0)^2
\right] }
\nonumber \\
&+&\frac{1}{2C_{a_3}} (n_3+Q_{v_3}^0)^2
-\sum_{i=1}^3 \frac{W_i}{2}
\nonumber \\
&+&\sum_{i=1}^2\frac{C_{D_i}}{4C_{a_i}C_{a_{i+1}}}
(n_i+Q_{v_i}^0)(n_{i+1}+Q_{v_{i+1}}^0).
\label{U0}
\end{eqnarray}
We consider a Coulomb blockade region
where the number of electrons can be controlled 
such that we can define a quantum state $|n_i\ra$ by the number of 
electrons $n_i$ in $i$-th FG as shown in Fig.~\ref{fig5}. 
The superposition state is constituted by the coupling 
region of  $|n_i\ra$ and  $|n_i\pm 1\ra$, which is 
realized when the parabolic charging energy of $|n_i\ra$ state
crosses that of $|n_i+ 1\ra$ state as discussed in Ref.~\cite{Makhlin}.
The crossover point where the charging energy of $n_i$ electrons 
equals that of the $n_i+1$ electrons is given by
\begin{equation}
(n_i+Q_{v_i}^0)^2 = (n_i+1+Q_{v_i}^0)^2.
\label{parabola}
\end{equation}
When we define an effective gate voltage $n_{G_i}$ given by
\begin{equation}
n_i+Q_{v_i}^0= n_{G_i} -1/2,
\end{equation}
then, the crossover point is given by $n_{G_i}=0$ and the superposition 
state is constituted around $n_{G_i} \ll 1$.
$n_{G_i}$ accounts for the effect of the gate voltage of the $i$-th cell.

Thus, the charging energy term as a function of $n_{G_i}$ is transformed to
\begin{equation}
U=\sum_{i=1}^3 h_i \sigma_i^z 
+\sum_{i=1}^2 J_{ij} \sigma_i^z \sigma_{i+1}^z  +{\rm Const}
\label{U},
\end{equation}
where 
\begin{eqnarray}
\!h_i \!&\!\!=\!\!& \!\! \frac{1}{2C_{a_i}} \!\!
\left[1\!+\!\!\frac{C_{D_i}^2}{C_{a_{i}}C_{a_{i+\!1}}} \!\right] \!n_{G_i}
\!+\!\!\frac{ C_{D_{i\!-\!1}} n_{G_{i-\!1}}\!+\!C_{D_i} n_{G_{i+\!1}}}{2C_{a_{i}}C_{a_{i+1}}}, \ \ \ \ \ 
\label{ngh} \\
\!h_3\!&\!\!=\!\!& \!\frac{1}{2C_{a_3}} n_{G_3} +\frac{C_{D_2}}{2C_{a_{2}}C_{a_{3}}} n_{G_2}, \\
\!J_{ij}\!&\!\!=\!\!&\!\frac{C_{D_i}}{4C_{a_i}C_{a_{j}}},  \label{JJ} 
\end{eqnarray}
and
\begin{eqnarray}
\sigma_i^x&=& |n_i \ra \la n_i+1 | + |n_i+1 \ra \la n_i|,
\\
\sigma_i^z&=&-|n_i \ra \la n_i | + |n_i+1 \ra \la n_i +1|,
\\
I_i&=&|n_i \ra \la n_i | +|n_i+1 \ra \la n_i +1|.
\end{eqnarray}
(Details of the definitions and derivations are noted in the Appendix).
We can define a height of charging energy $U_h$ 
as the coefficient of $n_{G_i}$ in Eq.(\ref{ngh}) and $U_w$ given by
\begin{eqnarray}
U_h &\equiv & \frac{1}{8C_{a_i}} \!\!
\left[1\!+\!\!\frac{C_{D_i}^2}{C_{a_{i}}C_{a_{i+\!1}}} \!\right] 
\end{eqnarray}
$U_w$ is defined by a voltage difference of the parabolas for $n_i$ and $n_i+1$.
When we see $Q_{v_i}^0$ as a function of the gate voltage $V_{\rm CG}$, 
we obtain $U_w=e/C_{A_i}$ from two equations of $n_i+Q_{v_i}^0(V_{\rm CG})=0$ and $n_i+1+Q_{v_i}^0(V_{\rm CG}+U_w)=0$.

When the quantum computations, such as Grover's algorithm~\cite{Grover}, 
are carried out, we have to maintain the electronic system 
in the range of its two-level system.
For that purpose, we have to maintain the superposition state between $|n_i\ra$ and 
$|n_{i+1}\ra$ in the present setup, and we have to take care 
such that $|n_{i-1}\ra$ or  $|n_{i+2}\ra$ states do not affect the target quantum operations,
when we apply the gate and substrate bias beyond $U_w$.
In contrast, the purpose of the quantum annealing process
is to finally obtain the $n_{G_i}=0$ state. 
Thus, as long as we can restore the point of  $n_{G_i}=0$, 
we can apply the gate and substrate bias beyond $U_w$.

\subsubsection{Numerical estimation of charging part}
\begin{figure}[h]
\centering
\includegraphics[width=8.0cm]{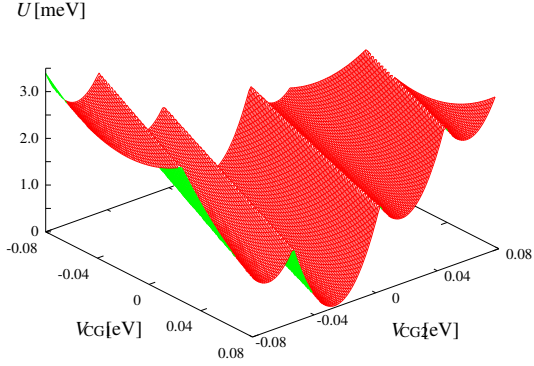}
\caption{
Numerical calculation of the charging energy Eq.~(\ref{U}) of three FGs
for $V_{\rm CG3}=V_{\rm sub}=V_{\rm CG1}$ and $V_{s_i}=0$ $(i=1,..,4)$. The electron density of the FG 
is $10^{20}$ cm${}^{-3}$. The size of FG is $L=W=10$ nm, the height of FG is $Z_{\rm FG}=100$ nm. The thickness of tunneling oxide is $d_{\rm ox}=3.5$ nm,  
and the thickness of the control gate oxide is $d_{\rm A}=8.2$ nm. The coupling ratio is 0.3. }
\label{fig6}
\end{figure}

\begin{figure}[h]
\centering
\includegraphics[width=8.0cm]{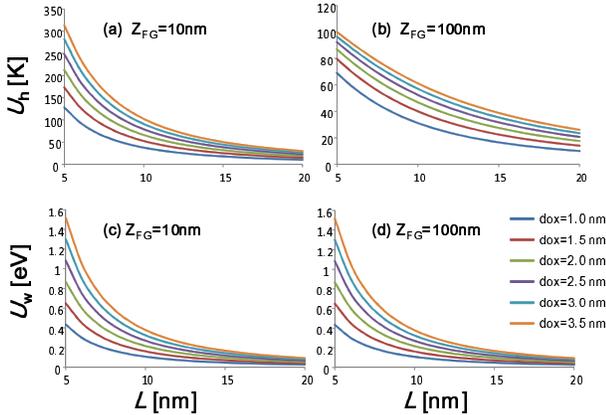}
\caption{
$U_{\rm h}$ and $U_{\rm w}$ that determine the single-electron effect (Fig.~2) are
calculated as a function of the size of FGs.
(a) $U_{\rm h}$ for $Z_{\rm FG}$=10 nm. 
(b) $U_{\rm h}$ for $Z_{\rm FG}$=100 nm.
(c) $U_{\rm w}$ for $Z_{\rm FG}$=10 nm. 
(d) $U_{\rm w}$ for $Z_{\rm FG}$=100 nm.
The thickness of the control gate oxide depends on $d_{\rm ox}$ by the coupling ratio $CR=0.3$. }
\label{figUhw}
\end{figure}

Figure \ref{fig6} shows an example of the numerical calculation of the charging energy
of three coupled FG cells for $CR=0.3$ and $L=15$ nm with $V_{\rm CG3}=V_{\rm CG1}$.
In order to experimentally observe single-electron effects, 
the charging energy should be sufficiently large to be observed at the operation temperature.
It is found that smaller $CR$ is better for increasing charging effects. 
Each parabola has its own electronic state of a fixed $n_i$.
Where two parabolas cross, the charging energies of different numbers of electrons 
cross as shown in Eq.~(\ref{parabola}).
Figure \ref{figUhw} shows $U_{\rm h}$ and $U_{\rm w}$
as functions of the size of FGs with various parameters of the thickness of the tunneling barrier $d_{\rm ox}$. 
$U_{\rm h}$ indicates a metric of charging energy, and therefore,  $U_{\rm h}$ 
is described as a unit of temperature  in Fig.~\ref{figUhw}.
$U_{\rm w}$ indicates a metric of gate controllability, and therefore, $U_{\rm w}$ is 
described as a unit of voltage in Fig.~\ref{figUhw}. 
From Figs.~\ref{figUhw}, it can be seen that a smaller FG is better for a larger charging energy as expected. 
It is also found that thicker $d_{\rm ox}$ is desirable.
Comparison of Fig.~\ref{figUhw}(a) and Fig.~\ref{figUhw}(b) reveals that
smaller height of FG is better for higher temperature operations.

\begin{figure}[h]
\centering
\includegraphics[width=8.0cm]{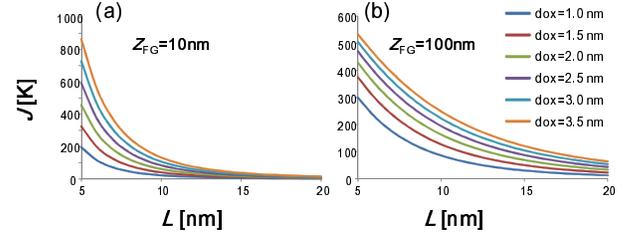}
\caption{
The coupling $J$, Eq.~(\ref{JJ}),  as a function of the size of FGs 
in which the width of FGs equals $L$. The vertical axis is described by a unit of temperature.
(a) The height of FG is 10 nm and (b) 100 nm.
The thickness of the control gate oxide depends on $d_{\rm ox}$ by the coupling ratio is $CR=0.3$. }
\label{figJ}
\end{figure}

Figure~\ref{figJ} shows $J$, Eq.~(\ref{JJ}),  as a function of the size of the FG.
As the size $L$ becomes larger, the magnitude of $J$ becomes smaller.
It can be also seen that 
the larger $d_{\rm ox}$ is better for larger $J$. This is because the larger $d_{\rm ox}$  
induces a larger area of Coulomb interactions with their neighboring FGs.

\begin{figure}[h]
\centering
\includegraphics[width=8.0cm]{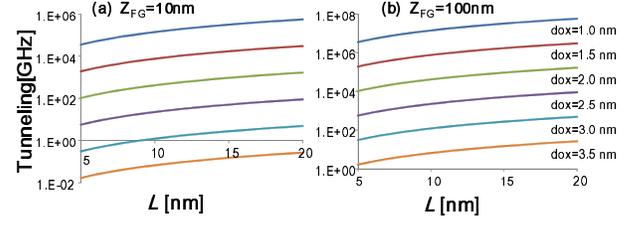}
\caption{
The magnitude of tunneling, Eq.(\ref{Tunnel}), as a function of the size of FGs.
(a) The height of FG is 10 nm. (b) The height of FG is 100 nm.  The coupling ratio is $CR=0.3$.
We use 1eV=2.41799$\times 10^5$ Hz.
}
\label{figTN1}
\end{figure}
\subsection{Derivation of tunneling part}
\subsubsection{Analytical formation of tunneling term}
Here, we derive the tunneling term in Eq.~(\ref{QAM}) by using the WKB approximation 
as shown in Ref.~\cite{Harrison}.
The tunneling term is expressed in the form of annihilation operators 
and given by
\begin{eqnarray}
\! \!
\lefteqn{ \sum_{k,k'} \Delta_{k,k'} c_{k_R}^\dagger c_{k_L} 
\approx 
v^2\! \! \int \frac{d^3kd^3k'}{(2\pi)^6}
 \frac{\hbar^2k_x}{2m_{\rm si}^* L } 
e^{ - \int k_v (x) dx }  \delta^{k_x}_{k_x'}
 c_{k^F_R}^\dagger c_{k^F_L}} \ \ 
\nonumber \\
%
%
%
%
&=&
N_LN_R  \frac{m_0R_y}{ m_{\rm si}^*} 
\left[\frac{\pi a_0 }{L}\right]^2  
e^{\left[ -\frac{d_{\rm ox}}{a_0} 
\sqrt{\frac{m_{\rm ox}^*(V_{\rm ox}-E_{\rm F}')}{m_0R_y} } \right]}
c_{k^R_F}^\dagger c_{k_F^L},
\nonumber \\
\label{Tunnel}
\end{eqnarray}
where $V_{\rm ox}$ is the potential height of the tunneling barrier, 
$m_{\rm si}^*=0.19m_0$ and $m_{\rm ox}^*=0.5m_0$ are the effective masses of electrons in Si and tunneling barrier, respectively
($m_0$is an electron mass in vacuum).
$c_{k_L}$ and  c$_{k_R}$ are annihilation operators of both sides of the tunneling barrier.
$k_F$ is a wave vector at a Fermi energy. $a_0\approx 0.0529$ nm is the Bohr radius and 
$R_y\approx 13.6$ eV is the Rydberg constant. 
$v$ is a volume of tunneling electrons, and
$N_L$, $N_R$ are the numbers of electrons of the two sides of the tunneling barrier, respectively,  
which participate in the tunneling event, 
given by 
\begin{eqnarray}
N_L =N_R =v\sum_k 1 = v \int dk_ydk_z k_{x_F} /(2\pi)^3.
\end{eqnarray}
This tunneling term is a function of the gate voltage $V_{\rm CG}$ through the 
shift of Fermi energy $E_F$ such as $E_{\rm F}'=E_{\rm F}+V_{\rm CG}$,
and $E_F$ is calculated by the FG doping concentration for which we use 10${}^{20}$ cm${}^{-3}$.
When $E_F'$ increases, the effective tunneling barrier is lowered and the tunneling rate 
increases (switch on). Conversely, when  $E_F'$ decreases, the effective tunneling barrier becomes larger, 
and the tunneling is switched off.
Thus, the tunneling can be switched on or off by controlling the gate and substrate bias.

\subsubsection{Numerical estimation of tunneling term}
Figure \ref{figTN1} shows a result of the numerical calculations of 
the tunneling term as a function of the FG sizes when 
$d_{\rm ox}$ and $Z_{\rm FG}$ are changed.
It can be seen that the tunneling rate increases as $L$ and $Z_{\rm FG}$ increase.
These are opposite trends to $U_{\rm h}$ and $U_{\rm w}$ in Fig.~\ref{figUhw}, 
and show that the increasing tunneling rate makes the saving of electrons difficult.
Tables 1 and 2 show typical results from Figs.~\ref{figUhw}-\ref{figTN1} as a summary 
of the basic calculations mentioned above.

Figure \ref{figTN2} shows the amplitude of the tunneling term as a function of 
$V_{\rm CG}$. It is seen that, when we change $V_{\rm CG}$ in the range of $\pm 1$ V,  
we can change the tunneling rate in the range of 10${}^4$. For example, 
for $Z_{\rm FG}=100$ nm and $d_{\rm ox}=3.5$ nm in Fig.~\ref{figTN2}(b),
the tunneling term for $V_{\rm CG}$=0 is approximately 26.4 Hz and can be neglected. 
That for $V_{\rm CG}$=-1 V corresponds to a case that an electron tunneling 
is expected to occur by approximately 2.2kHz.

\subsubsection{Normally-off and normally-on} 
\begin{figure}[h]
\centering
\includegraphics[width=8.0cm]{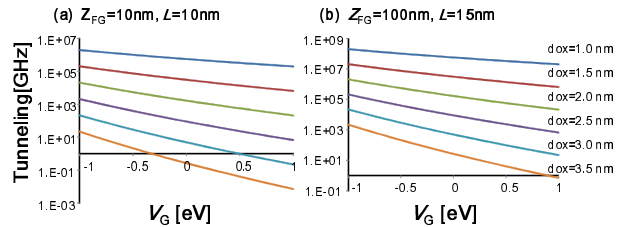}
\caption{
The magnitude of tunneling, Eq.(\ref{Tunnel}), as a function of the gate voltage $V_{\rm CG}$.
(a) The height of FG is 10 nm. (b) The height of FG is 100 nm.  The coupling ratio is $CR=0.3$. }
\label{figTN2}
\end{figure}

So far, we have considered a QAM in which tunneling is switched on 
when external biases are applied.  Thus, these devices can be called 
``normally-off'' devices.
On the other hand, as $d_{\rm ox}$ becomes thinner, 
the tunneling frequency becomes larger and larger 
and the tunneling is always on even for $V_{\rm CG}=0$.
For example in Fig.~\ref{figTN2}(a), the tunneling term for $d_{\rm ox}=1.5$ nm of $L=15$ nm is 17.1 THz, 
and in Fig.~\ref{figTN2}(b), the tunneling term for $d_{\rm ox}=1.5$ nm is 1717 THz.
For these cases, we can switch off the tunneling by applying positive electric voltages. 
Thus, we call these devices ``normally-on" devices.
Compared with normally-off devices, the power consumption of 
the normally-on devices is considered to be larger. 
Thus, we mainly consider the normally-off devices and discuss the normally-on 
devices briefly in the Appendix.

\begin{table}[t]
{TABLE 1. Examples of the result of the analytical calculations  for $d_{\rm ox}=2.5$ nm and $Z_{\rm FG}=$10 nm.} 
\begin{tabular}{l  | r r c | r}
\hline\hline
Size         & \ \ \ $J$ [K]  & \ \ \ $U_{\rm h}$ [K]  & \ \ \ Tunnel [GHz]  &\ \ \  $U_{\rm w}$ [eV] \\
\hline
$L$=5 nm  & 596.0   & 249.7          & 5.61      & 1.08      \\
$L$=10 nm  & 85.8  & 80.4             & 22.5     & 0.27       \\
$L$=15 nm &   23.5   & 39.5           & 50.5   & 0.12      \\
\hline
\end{tabular}
\end{table}
\begin{table}[t]
{TABLE 2. Examples of the result of the analytical calculations  for $d_{\rm ox}=3.5$ nm and 
$Z_{\rm FG}=100$ nm.} 
\begin{tabular}{l  | r r c | r}
\hline\hline
Size         &\ \ \ $J$ [K] & \ \ \ $U_{\rm h}$ [K] &  \ \ \ Tunnel [GHz] &\ \ \ $U_{\rm w}$ [eV] \\
\hline
$L$=5 nm  &  534.4   & 100.1       & 1.65     & 1.52 \\
$L$=10 nm &  251.0  & 61.5         & 6.61    & 0.38      \\
$L$=15 nm &  124.0   & 39.3        & 14.9     & 0.17     \\
\hline
\end{tabular}
\end{table}

\section{Results of TCAD simulations}\label{sec_tcad}
\begin{figure}
\centering
\includegraphics[width=8.6cm]{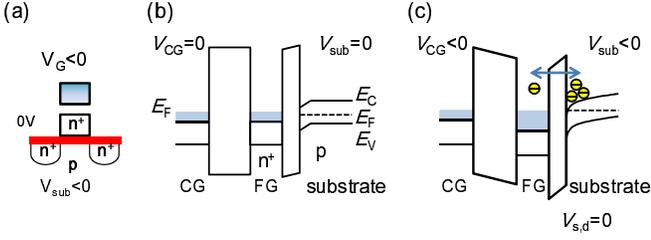}
\caption{
Schematics of FG cell in the proposed QAM.
(a) Doping pattern of an FG cell in our QAM. 
The FG and substrate consist of n-type and p-type semiconductors, respectively.
(b) Band structure of no bias and no tunneling events.
(c) Band structure of finite tunneling. Both the gate bias and substrate bias are applied negatively.
}
\label{fig_dope}
\end{figure}

In general, FG memories consist of many different regions with different 
doping concentrations and doping types. 
Different regions have different work functions without bias.
Typically, a p-type substrate,  n-type source/drain and an n-type FG are employed
where the doping concentration of the FG is higher than that in the substrate. 
Figure \ref{fig_dope}(a) shows the typical doping pattern.
Figures \ref{fig_dope}(b) and (c) show the band structures depending on a bias voltage.  

Here, we would like to show the results of the conventional TCAD simulation in our targeted parameter regions.
We calculated the device characteristics of the three FG array of Fig.~\ref{fig_dope}(a) 
by using our in-house TCAD simulator.
This simulator was developed for conventional 
semiconductor devices, and cannot include the single-electron effect.
The number of the coupled cells is limited by our calculation resources. 
Here we do not calculate $h_j$ (Eq.(\ref{ngh})) nor $J_{ij}$ (Eq.(\ref{JJ})), 
because capacitances change depending on applied biases as a result of the change of  depletion layers.
Elaborate estimation will be required.
Accordingly, the crossover point of $n_{\rm G}\approx 0$ in Fig.~\ref{fig5} should be 
estimated in the near future.

\begin{figure}
\centering
\includegraphics[width=8cm]{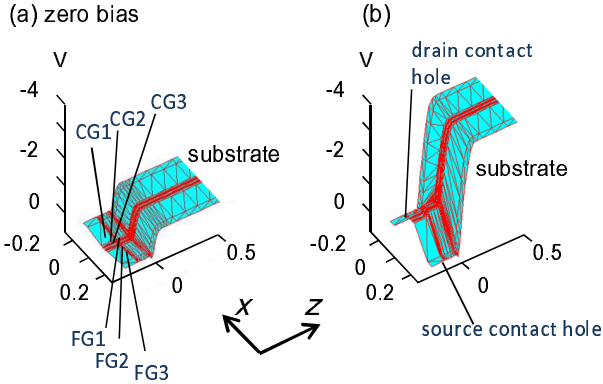}
\caption{
Bird's-eye views of the electrical potential profile of three FG cells
calculated by  in-house TCAD simulator. 
The interface between the tunneling barrier and 
substrate is $z=0$ where,  for simplicity, the tunneling barriers are not shown.
The source and drain in the substrate are set close to the outer two FGs, 
and their contact holes exist at $|x| \ge 0.2075 \mu$m and $z<0$. 
The thickness of the substrate is 1 $\mu$m, and only the upper part of the substrate 
is shown. 
$L=W=15$ nm, $d_{\rm ox}=3.0$ nm and $d_{\rm A}=7.0$ nm.
($CR=0.3$ in Eq.(\ref{CR})). 
The heights of FG and CG are 0.1 $\mu$m.
The doping rates of FG, source, and substrate are given by 
10$^{20}$ cm${}^{-3}$, 2.0$\times$10$^{19}$ cm${}^{-3}$ and 3$\times$10$^{17}$ cm${}^{-3}$, respectively.
(a) $V_{\rm CG1}=V_{\rm CG2}=V_{\rm CG3}=V_{\rm sub}=0$. (b) $V_{\rm sub}=-3$ V, and
$V_{\rm CG1}=V_{\rm CG2}=V_{\rm CG3}=-1$ V.
}
\label{figtcad}
\end{figure}

\begin{figure}
\centering
\includegraphics[width=8cm]{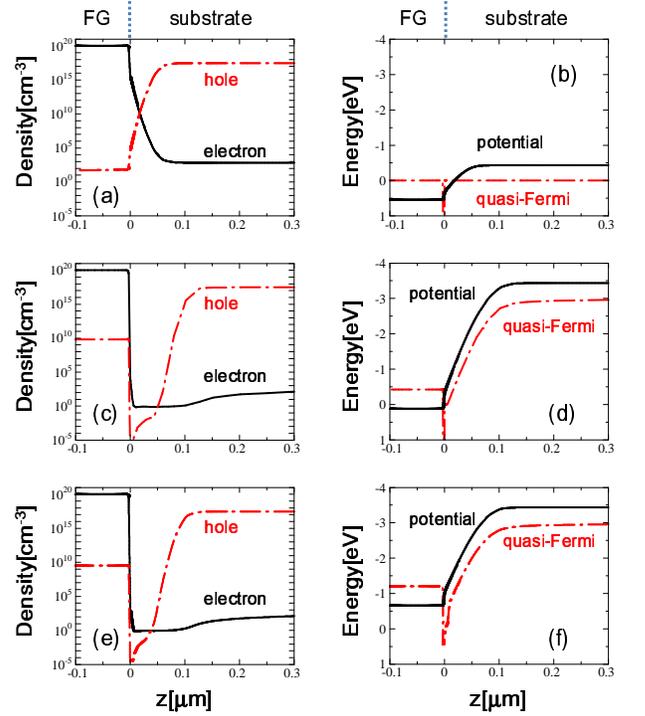}
\caption{
The carrier densities (electrons and holes), the potential energies, 
and the quasi-Fermi energies of the center FG2, as a function of the distance
from the surface of the substrate.
(a) and (b) show the results of zero bias ($V_{\rm CG1}=V_{\rm CG2}=V_{\rm CG3}=V_{\rm sub}=0$).
(c) and (d) show the results of $V_{\rm CG1}=V_{\rm CG2}=V_{\rm CG3}=0$ V and $V_{\rm sub}=-3$ V.
(e) and (f) show the results of $V_{\rm CG1}=V_{\rm CG2}=V_{\rm CG3}=-1$ V and $V_{\rm sub}=-3$ V.
Those of the FG1 and FG3 have similar values as these results.}
\label{figdensity}
\end{figure}

In order to generate electrons at the surface of the substrate (Fig.~\ref{fig_dope}(b) and (c)), 
$V_{\rm CG} > V_{\rm sub}$ is required~\cite{Grove}.
Figure \ref{figtcad} shows the bird's eye views of the electronic potentials 
of zero bias (Fig.~\ref{figtcad}(a)) and the case of 
$V_{\rm sub}=-3$ V and
$V_{\rm CG1}=V_{\rm CG2}=V_{\rm CG3}=-1$ V (Fig.~\ref{figtcad}(b)).
Figure~\ref{figdensity} shows the calculated electron and hole densities, potential energies and quasi-Fermi energies.
It is realized that the electron density is larger than the hole density at the substrate interface.
It can be also seen that Figs.~\ref{figdensity} (c) and (d) are similar to Figs.~\ref{figdensity}(e) and (f), respectively.
This means that the case of $V_{\rm CG1}<0, V_{\rm CG2}<0$, and $V_{\rm CG3}<0$ is an extension of the case of
$V_{\rm CG1}=V_{\rm CG2}=V_{\rm CG3}=0$ when $V_{\rm sub}<0$. 
The quasi-Fermi energy and potential energy of the FGs in Fig.~\ref{figdensity}(f) become higher
than those of Fig.~\ref{figdensity}(d) while the bottom of the quasi-Fermi energy 
at the substrate surface does not significantly change. This is the expected result in Fig.~\ref{figConcept}.

\begin{figure}
\centering
\includegraphics[width=8cm]{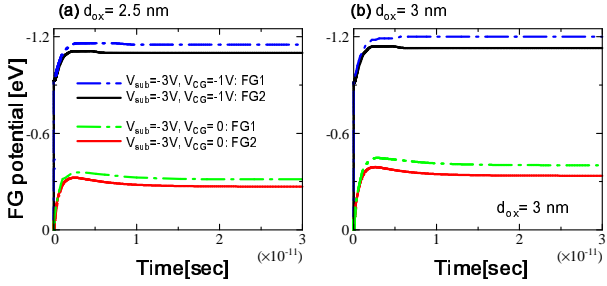}
\caption{
Time dependent analysis of the center FG (FG2) potentials calculated by our in-house simulator.
(a) $d_{\rm ox}=2.5$ nm and (b) $d_{\rm ox}=3.0$ nm.
Other two FGs (FG1 and FG3) show similar behaviors.
Slight shift of the potentials can be seen when these results 
are compared with those of Fig.~\ref{figdensity}. 
This is a problem of our simulator in which the different calculations 
are required when time-dependent analysis is carried out. 
}
\label{figtime}
\end{figure}

We also calculated a time dependent FG potential as shown in Fig.~\ref{figtime}
to see the dynamic behavior of tunneling.
We apply finite voltages on the gate and substrate at $t=10^{-15}$ sec, starting from zero bias state.
It can be seen that the time scale of the change of the potentials is of the order of $10^{-11}$ sec.
Thus, tunneling event is very fast as estimated in the previous sections.
%

\section{Noise and decoherence}\label{sec_decoherence}
Here, we discuss the noise and decoherence problems in the QAM based on an FG array.
In the conventional FG array, the thickness of the tunneling oxide 
is of the order of 7 nm to store the charge in the FG for many years.
In the present setups, we apply a thinner tunneling oxide ($d_{\rm ox} \le 3.5$ nm ) to control 
switching on and off the tunneling. 
Thus, natural leakage of charges occurs more often than in the case of the conventional FG memory.
Therefore, after the QAM process, the results should be quickly transferred 
to a conventional memory block.

The unexpected trap sites are ones of the problems of 
NAND flash memories~\cite{Paolucchi1,Paolucchi2,Aritome2}.  
These will become more serious when we use an FG array as a qubit system.
However, the physical properties of the traps persist on each cell
as shown in Ref.~\cite{Kim,Chen}.
Once we check the trap distributions and their properties and store that 
information to conventional memory,
we could apply appropriate voltages depending on each FG cell  and 
improve the QAM performance.

Next, let us estimate the coherence time by taking into account the electron-phonon interactions,
following the discussion in Ref.~\cite{tana0}.
We assume that the acoustic phonon in the tunneling material of SiO${}_2$ is 
the principal origin of decoherence and apply the standard spin-boson theory~\cite{Legget}.
According to Leggett {\it et al.}~\cite{Legget}, 
the complete information about the effect of the environment is encapsulated 
by the spectral function ${\mathcal J}(\omega)$. In the present case, the two-state system 
is formed through the tunneling of SiO${}_2$ and it  has two types of 
spectral function given by~\cite{Garcia,Wurger}
\begin{equation}
{\mathcal J}(\omega)= \frac{\gamma^2}{\pi \rho c^5}\omega^3+\frac{\gamma^2 \nu^2}{\rho c^3 d^2}\omega.
\label{spectral}
\end{equation}  
The first term shows a superohmic dissipation and corresponds to 
an underdamped coherent oscillation, whose 
damping rate $\Gamma_{\rm so}$ at $T=0$ is given by~\cite{tana0}
\begin{equation}
\Gamma_{so}=\frac{\gamma^2\tilde{\Delta}^3}{4\pi \hbar \rho c^5},
\end{equation}
where $\tilde{\Delta}$ is the renormalized tunneling frequency $\Delta$, 
and given by
\begin{eqnarray}
\tilde{\Delta}
&=&\Delta \exp \left(
-\frac{\gamma^2\omega_c^2}{2\pi^2 \hbar \rho c^5}
\right).
\end{eqnarray}
When we use $\gamma \sim 10$ eV, $c\sim 4300$ m/s, $\rho \sim 2200$ kg/m${}^{-3}$,
and $\omega_c=k_B \Theta_D /\hbar$ with 
$\Theta \sim 450$K, we have $\tilde{\Delta}\sim \Delta e^{-1323}$, 
and thus this term can be neglected.
The second term of Eq.~(\ref{spectral}) expresses an ohmic dissipation. 
The parameter $\alpha$ of the ohmic dissipation is given by $\alpha=\gamma^2 \nu^2/(2\pi^2\hbar \rho c^3d^2) \sim 7.05 \times 10^{-9}$.
Then $P(t)\equiv \la \sigma_z (t) \ra=P_{\rm coh} (t) +P_{\rm inc}(t)$ is given by
\begin{eqnarray}
P_{\rm coh}(t) &\approx& \cos \Delta t \exp \left\{ -\frac{\pi\alpha}{2}\Delta t\right\}, \\
P_{\rm inc}(t) &\approx & \alpha \{ \Delta t ( ci(\Delta t) \sin \Delta t -si(\Delta t) \cos (\Delta t) \},\ \ 
\end{eqnarray} 
where $ci(y)=-\int_y^\infty (\cos x)/x dx$ and $si(y)=-\int_y^\infty (\sin x) /x dx$. 
When we estimate the coherence time from the exponential part of $P_{\rm coh}(t)$
such as  $\alpha \pi \Delta t_{\rm coh} /2=1$, we have $t_{\rm coh}\sim 4.33$ msec for $\Delta=10$K
and $t_{\rm coh}\sim 0.433$ msec for $\Delta=100$K, respectively. 
Therefore, the increase of the tunneling excites more phonons, and a low tunneling rate is desirable. 
However, tunneling that is too slow induces many noises~{\cite{Paolucchi1,Paolucchi2,Aritome2}. 
Thus, there will be an optimal point for the tunneling rate, depending on each system.

\section{Discussion}\label{sec:discuss}
In Ref.~\cite{tana0,tana1}, we have proposed a charge-qubit system~\cite{NEC,transmon} 
based on coupled quantum dots (CQDs)~\cite{Stockklauser}.
When we regard FGs as quantum dots (QDs),
the previous CQD proposal is different from the present proposal in that there are additional FGs  
stacked on the first FG layer~\cite{tana2}. 
Although additional fabrication processes are required for the CQD system, 
the structure of the CQD system can be manufactured 
by the same set of photo masks as that in the present proposal.
Because the electron moves in the close-coupled QDs in the previous proposal, 
its coherence time is considered to be longer than 
that of the present proposal. 
Besides the fact that the present proposal is more close to the conventional flash memory, 
the advantage of the present proposal is that the electron tunneling 
is directly changed by the external electrodes.
In contrast, the tunneling barrier of the CQD system is fixed, and qubit states are mainly controlled in the 
rotating frame by applying high-frequency electric field.

We have to readout the final state after the annealing process.
At the readout stage,  the number of electrons in the FG should be detected
by applying gate voltage and switching on current between the source and drain.
As in the conventional FG array, the applied voltage of the reading out 
is smaller than that of the `switching-on' process.
The readout process should not change the number of electrons in the FG array.
The thickness of the tunneling oxide of the normally-off type is greater than
that of the normally-on type (as discussed in the Appendix).  
Thus, it is considered that the normally-off  
type is more easily constructed than the normally-on type. 
In any event, detailed optimizations of the structure and operation voltages are 
subjects for future work.

\section{Conclusion}\label{sec:conclusion}
We have shown that the conventional FG memory structure can be used 
as the quantum annealing machine.
It is assumed that the FG is sufficiently small for 
the appearance of the single-electron effect.
The electrostatic Coulomb interaction between FG cells is 
the origin of the Ising interaction.
The tunneling between the FG and substrate is brought about 
by applying voltages to both the CG and substrate,
while the source/drain voltage is set to zero.
Using TCAD simulation, we have shown an example of the device characteristics 
under possible bias conditions. 
Optimization of many parameters is a subject for future work.

\acknowledgements
 We thank A. Nishiyama, M. Koyama, S. Yasuda, H. Goto, D. Matsushita, K. Matsuzawa and T. Ishihara for discussions.

\appendix
\section{Detailed derivation process of the charging energy}
Here, we show additional information for the derivation process of Eq.(\ref{U}).
The parameters in Eq.(\ref{U0}) in the text are given by
\begin{eqnarray}
C_{a_1}&\!\!=\!\!& C_{A_1}+C_{B_1}+C_{D_1}+C_{F_1}+C_{H_1}+C_{I_1}, 
\nonumber \\
C_{a_2}&\!\!=\!\!& C_{A_2}+C_{B_2}+C_{D_2}+C_{F_2}+C_{H_2}+C_{I_2},
\nonumber \\
&\!\!+\!\!&C_{D_1}+C_{E_1} -\frac{C_{D_1}^2}{C_{a_1}},
\nonumber \\
C_{a_3}&\!\!=\!\!& C_{A_3}+C_{B_3}+C_{D_3}+C_{F_3}+C_{H_3}+C_{I_3},
\nonumber \\
&\!\!+\!\!&C_{D_2}+C_{E_2} -\frac{C_{D_2}^2}{C_{a_2}},
\nonumber \\
Q_{v_1}^0&\!\!=\!\!& C_{A_1}V_{\rm CG1}+C_{B_1}V_{\rm sub}+C_{F_1}V_{\rm CG2}+C_{H_1}V_{s_1}+C_{I_1}V_{d_1},
\nonumber \\
Q_{v_2}^0&\!\!=\!\!& C_{A_2}V_{\rm CG2}+C_{B_2}V_{\rm sub}+C_{F_2}V_{\rm CG3}+C_{H_2}V_{s_2}+C_{I_2}V_{d_2},
\nonumber \\
Q_{v_3}^0&\!\!=\!\!& C_{A_3}V_{\rm CG3}+C_{B_3}V_{\rm sub}+C_{H_3}V_{s_3}+C_{I_3}V_{d_3},
\nonumber \\
W_{1} &\!\!=\!\!& C_{A_1}V_{\rm CG1}^2+C_{B_1}V_{\rm sub}^2+C_{F_1}V_{\rm CG2}^2+C_{H_1}V_{s_1}^2+C_{I_1}V_{d_1}^2,
\nonumber \\
W_{2} &\!\!=\!\!& C_{A_2}V_{\rm CG2}^2+C_{B_2}V_{\rm sub}^2+C_{F_2}V_{\rm CG3}^2+C_{H_2}V_{s_2}^2+C_{I_2}V_{d_2}^2,
\nonumber \\
W_{3} &\!\!=\!\!& C_{A_3}V_{\rm CG3}^2+C_{B_3}V_{\rm sub}^2+C_{H_3}V_{s_3}^2+C_{I_3}V_{d_3}^2, \!
\end{eqnarray}
where $V_{d_i}=V_{s_{i+1}}$.
Following Ref~.\cite{Makhlin}, we consider the region around
\begin{equation}
(n_i+Q_{v_i}^0)^2 = (n_i+1+Q_{v_i}^0)^2
\label{parabola_cond}
\end{equation}
This is the region where the charging energy of $n_i$ electrons 
equals  that of the $n_i+1$ electrons, and 
the $|n_i\ra$ and $|n_i+1\ra$ states constitute a superposition state. 
We use the effective gate voltage $n_{G_i}$ given by
\begin{equation}
n_i+Q_{v_i}^0= n_{G_i} -1/2.
\end{equation}
($n_{G_i} \ll 1$).
For this region, we can approximate the following equation
\begin{eqnarray}
\sum_{m=0}^{1}(n_i+m+Q_{v_i}^0)^2 & \rightarrow &
\frac{1}{2}  n_{G_i} \sigma_i^z +\left(n_{G_i}^2 +\frac{1}{4}\right) I_i, \ \ \ \ \ 
\end{eqnarray}
where $\sigma^z$, $I_i$ are Pauli matrix and unit matrix, respectively,  
based on $\{|n_i+1 \ra,|n_i \ra \}$ system.
Thus, we obtain Eq.(\ref{U}) and
\begin{eqnarray}
\lefteqn{ {\rm Const}=\sum_{i=1}^2 \left[ 
\frac{1}{2C_{a_i}} 
\left(1+\frac{C_{D_i}^2}{C_{a_{i}}C_{a_{i+1}}} \right) 
\left( n_{G_i}^2+\frac{1}{4} I_i \right)
-\frac{W_i}{2}
\right] }\nonumber \\
&+& \frac{1}{2C_{a_3}}\left(n_{G_3}^2+\frac{1}{4} \right) I_3-\frac{W_3}{2}
+\sum_{i=1}^2\frac{C_{D_i}}{C_{a_i}C_{a_{i+1}}} I_i I_{i+1}. \ \ \ \ \ \ \
\end{eqnarray}
\begin{figure}
\centering
\includegraphics[width=8.6cm]{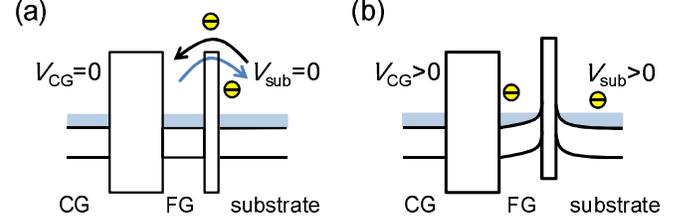}
\caption{
Schematics of FG cell of a normally-on QAM.
(a) Band structure without bias.  Tunneling is switched on.
(b) Band structure without tunneling events. Both the gate bias and substrate bias are applied positively.
}
\label{normally-on}
\end{figure}

\section{Normally-on and normally-off}
In the main text, we consider a case in which tunneling occurs 
only when the gate and substrate biases are applied.
These devices can be called "normally-off" devices. 
In this section, we briefly discuss a ``normally-on'' device
in which tunneling is switched off only 
when the gate and substrate bias are applied. 

Figure~\ref{normally-on} shows an example of 
the normally-on set up.
The electron tunneling always occurs between the FG and the substrate without bias as shown in Fig.~\ref{normally-on}(a).
The effective thickness of the effective tunneling barrier is changed by the 
change of the potential of the FG and substrate.
If we prepare a p-type substrate like that of the conventional NAND flash memory,
the current from the substrate to the source and drain flows when 
the tunneling is switched off (Fig.\ref{normally-on}(b)).  
This is because the source and drain are n-type semiconductors. 
Thus, in the normally-on type QAM, there are always some current flows in the cell.
Detailed calculations are subjects for future work.


\end{document}